\documentclass[reprint,amsmath,amssymb,twocolumn]{revtex4-2}
\usepackage[colorlinks,citecolor=blue,linkcolor=blue,anchorcolor=blue,filecolor=blue,
urlcolor=blue]{hyperref}
\usepackage{graphicx}

\begin{document}

\title{Surface and curvature tensions of relativistic models}

\author{Mariana Dutra$^{1}$, Odilon Louren\c{c}o$^{1}$ and D\'ebora P. Menezes$^2$}

\affiliation{
$^1$Departamento de F\'isica e Laborat\'orio de Computa\c c\~ao Cient\'ifica Avan\c cada e Modelamento (Lab-CCAM), Instituto Tecnol\'ogico de Aeron\'autica, DCTA, 12228-900, S\~ao Jos\'e dos Campos, SP, Brazil \\ 
\mbox{$^2$ Depto de Física, CFM, Universidade Federal de Santa Catarina, Florianópolis, SC, CP:476, CEP 88.040-900, Brazil }
}

\date{\today}

\begin{abstract}
In the present paper, we show a simple method to obtain fittings for the surface and curvature tensions. The method uses
the nuclear mass of a spherical fully ionized atom and a simple expression for the binding energy such that a least square fit is found when confronted with the Atomic Mass Evaluation (AME) 2020. The fittings are then used to evaluate the pasta phase free energy per particle, which is confronted with the one obtained with a Thomas-Fermi fitting. The results are very encouraging and suggest that this recipe can be safely used whenever the surface and curvature tensions are necessary. 
\end{abstract}

\maketitle

\section{Introduction}

Neutron stars (NSs) are exotic compact objects whose constitution has been a source of intense investigations over the last decades. While their cores can be made of hadronic matter, quark matter, or perhaps a mixture of both, their crusts are believed to contain outer and inner parts. For a review, interested readers can rely on refs. \cite{livro-Glendenning:2012,livro-Weber:2023,Debora2021-267}. 

In the present work, we discuss only the pasta phase, probably present in the inner crust, and the importance of a reliable prescription for the surface and curvature tensions compatible with the model used to describe the neutron star core.  The pasta phase is constituted by nonspherical complex structures that appear due to frustration in sub-saturation nuclear densities~\cite{PhysRevLett.50.2066,hashimoto84}. Although expected to be present only in a small range of densities and temperatures, the pasta phase can probably leave signatures in different astrophysical phenomena
~\cite{soft,Okamoto:2013tja, PhysRevC.70.065806, Lin:2020nxy,latetime}. Moreover, neutrino diffusion is probably affected by the pasta phase in protoneutron stars~\cite{PhysRevC.83.035803}.

While in NSs, the system obeys charge neutrality and $\beta$-equilibrium conditions, the same kind of pasta structure is expected to appear during the supernova core-collapse stage, but in this case, the temperature is higher and the proton fraction is fixed~\cite{ravenhall1983,helena2012}. 

The degree of complexity of the pasta phase is very model-dependent and the literature shows calculations that foresee 
1D, 2D, and 3D geometries in a single unit cell - as in the original references already mentioned, density fluctuations that allow coexistence of these geometries ~\cite{flut1,Pelicer:2021ils} and even non-trivial structures resembling waffle, parking garage and triply periodic minimal surfaces (TPMS)~\cite{complex1,complex2,complex3}.

The pasta phase size decreases as temperature increases and may be a very thin layer between two homogeneous phases at certain temperatures~\cite{oldpasta,PRC82, PhysRevC.85.059904-erratum-PRC82.055807,PRC85,PhysRevC.105.025806,PhysRevD.106.063020,PhysRevC.103.055810,PhysRevC.102.015806}. The crust-core transition density can be obtained in different ways~\cite{PhysRevC.79.035804,PRC82,EurPhysJA.50.44,Chatterjee_2019}, but generally depends on equations of state (EOS) parameterized to satisfy nuclear matter bulk properties. One common characteristic of the crust-core transition density is its dependence on the surface tension. 

A simple prescription used to calculate the pasta phase is the coexistence phase approximation (CPA)~\cite{PhysRevC.72.015802,oldpasta} but it depends on the surface tension expression obtained from a more sophisticated numerical method, the Thomas-Fermi (TF) approximation~\cite{PRC85}. But, what if this TF fitting is not available for a certain relativistic model? Any other prescription used so far is not as consistent. Hence, this is the main reason for the present paper: to provide reliable and consistent surface and curvature expressions for different relativistic models. 

To tackle this problem, we use the nuclear mass of a spherical fully ionized atom, as proposed in~\cite{Hoa_2021} and a simple expression for the binding energy such that a least square fit is found when confronted with the Atomic Mass Evaluation (AME) 2020 \cite{AME2020}. The details are given in Section \ref{AME2020}.
For each relativistic hadronic model chosen, a mean field approximation (RMF) is performed and the nuclear matter properties are obtained, including the binding energy. The procedure is standard in the literature and can be seen in detail, for instance,  in~\cite{livro-Glendenning:2012,PhysRevC.90.055203}. 

As for the chosen parametrizations of the RMF model, we have opted for some previously selected ones in Ref.~\cite{debora2019}: \mbox{G2$^*$}, IUFSU, \mbox{DD-F}, TW99, and \mbox{DD-ME$\delta$}. They were shown to be consistent with observational data from LIGO/Virgo Collaboration~\cite{ligo} on the tidal deformabilities of the GW170817 event, as well as being capable of producing massive stars. We also consider the following RMF parametrizations studied in Ref.~\cite{brett23}: BSR1, BSR2, BSR3, BRS4, BSR8, BSR9, BSR10, BSR15, BSR16, BSR17, FSUGZ00, FSUGZ03, FSUGZ06 and IUFSU*, all of them in agreement with macroscopic properties of neutron stars, and also with data related to giant monopole resonances, charge radii, and ground state binding energies of some spherical nuclei, namely, $^{16}\rm O$, $^{34}\rm Si$, $^{40}\rm Ca$, $^{48}\rm Ca$, $^{52}\rm Ca$, $^{54}\rm Ca$, $^{48}\rm Ni$, $^{56}\rm Ni$, $^{78}\rm Ni$, $^{90}\rm Zr$, $^{100}\rm Sn$, $^{132}\rm Sn$, and $^{208}\rm Pb$. They are also compatible with the constraint deduced from the analysis of the excitation energy of the isobaric analog state, based on Skyrme-Hartree-Fock calculations~\cite{danielewicz2014}, along with data from neutron skin thickness of $^{208}\rm Pb$. For the sake of comparison, we also include other ``popular'' parametrizations in our analysis: NL3 and NL3$\omega\rho$.

To guarantee that the obtained fitting is reasonable, the pasta phase computed with the TF surface tension is confronted with the one generated by the present fitting for some models. They are indeed close and the transition density from the pasta to the homogeneous phase is coincident.

Of course, anyone interested in using a different model, can follow the prescription given in the present paper and obtain the desired fitting for the surface and curvature tensions. 

\section{AME2020 Fitting results}
\label{AME2020}

The nuclear mass of a spherical fully ionized atom can be expressed as~\cite{Hoa_2021}
\begin{align}
&M(A,Z) = Zm_p + (A-Z)m_n + A\frac{\epsilon_B(n_{\rm{eq}},y)}{n_{\rm{eq}}} 
\nonumber \\
&+ 4\pi r_{\rm eq}^2\left[\sigma_s(y,T=0) + \frac{2\sigma_c(y,T=0)}{r_{\rm eq}}\right] 
+ \frac{3}{5}\frac{e^2Z^2}{r_{\rm eq}},
\label{eq:mass}
\end{align}
where $Z$ is the proton number, $A$ is the mass number, and $y = Z/A$ is the proton fraction. $m_p$ ($m_n$) is the proton (neutron) mass, $r_{\rm eq} = (4\pi n_{\rm{eq}}/3)^{-1/3}A^{1/3}$ is the nuclear radius, and $n_{\rm{eq}}$ is the equilibrium density of infinite nuclear matter determined from the condition given by
\begin{align}
\frac{\partial(\epsilon_B/n)}{\partial n}\Big|_{y,n=n_{\rm eq}} = 0.
\end{align}
The surface and curvature tensions at zero temperature in Eq.~(\ref{eq:mass}) can be written, respectively, by~\cite{Ravenhall_1983,Hoa_2021}
\begin{align}
\sigma_s(y,0) = \sigma_0 \frac{2^{(p+1)} + b_s}{y^{-p} + b_s + (1-y)^{-p}}, 
\label{eq:sigmast0}
\end{align}
and
\begin{align}
\sigma_c(y,0) = 5.5 \sigma_s(y,0) \frac{\sigma_{0,c}}{\sigma_0} (\beta - y),
\label{eq:sigmact0}
\end{align}
where $\sigma_0, b_s, \sigma_{0,c}, \beta$ are the parameters to be adjusted. Here we use $p=3$. Finally, the binding energy per nucleon can be calculated as follows
\begin{align}
B_{\rm theo} = \frac{1}{A}[Z m_p + (A-Z)m_n - M(A,Z)].
\label{eq:bind_energy-exp}
\end{align}

For each model/parametrization, we determine the parameters  $\sigma_0, b_s, \sigma_{0,c}, \beta$ from a least squared fit of Eq.~(\ref{eq:bind_energy-exp}) to Atomic Mass Evaluation (AME) 2020 \cite{AME2020}. For the fitting procedure, we have used the experimental binding energy values,~$B_{\rm exp}$,  related to nuclei in which $N \ge 8$ and $Z \ge 8$. We also use the same uncertainty for all data. In order to evaluate the quality of the fitting, we also calculated the $\chi^2$ value from
\begin{align}
\chi^2 = \frac{1}{N_{\rm{d}} - 4}\sum_{i=1}^{N_{\rm{d}}}
\frac{[B^i_{\rm theo} - B^i_{\rm exp}]^2}{\Delta B_i^2},
\label{eq:chi2}
\end{align}
where $N_{\rm{d}}$ is the number of data points. As the experimental error is negligible, $\Delta B_i$ has only theoretical components. We estimate this error as the same for all nuclei given by $\Delta B_i=0.04$~MeV. The parameters found by such a method are presented in Table~\ref{tab:fit}.
\begin{table*}[!htb]
\tabcolsep=0.2cm
\centering
\caption{Optimised surface and curvature parameters, with their respective uncertainties, for different parametrizations of the RMF model fitted from the AME2020~\cite{AME2020} data. The critical temperature for symmetric nuclear matter case, $T_c^{\rm SM}$, is also furnished.}
\begin{tabular}{l|c|c|c|c|c|c}
\hline
Models & $\sigma_0$ & $b_s$ & $\sigma_{0,c}$ & $\beta$ & $\chi^2$ & $T_c^{\rm SM}$\\
    & (MeV.fm$^{-2}$) &  & (MeV.fm$^{-1}$) &     &      & (MeV)\\ 
\hline
BSR1~\cite{bsr} & $1.04475 \pm 0.00093$ & $24.74119 \pm 0.24696$ & $0.10236 \pm 0.00312$ & $0.57178 \pm 0.00439$ & $0.73636$ & $13.910$ \\
BSR2~\cite{bsr} & $1.04672 \pm 0.00095$ & $22.65076 \pm 0.22719$ & $0.10661 \pm 0.00318$ & $0.57923 \pm 0.00439$ & $0.75192$ & $13.809$\\
BSR3~\cite{bsr} & $1.06758 \pm 0.00096$ & $18.19746 \pm 0.18153$ & $0.11194 \pm 0.00321$ & $0.55727 \pm 0.00394$ & $0.75376$ & $13.855$\\
BSR4~\cite{bsr} & $1.06173 \pm 0.00099$ & $16.68266 \pm 0.17282$ & $0.11818 \pm 0.00331$ & $0.57924 \pm 0.00414$ & $0.79438$ & $13.735$\\
BSR8~\cite{bsr} & $1.05532 \pm 0.00090$ & $24.78982 \pm 0.23951$ & $0.09716 \pm 0.00303$ & $0.53254 \pm 0.00404$ & $0.69886$ & $14.168$\\
BSR9~\cite{bsr} & $1.07098 \pm 0.00090$ & $22.75487 \pm 0.21611$ & $0.09889 \pm 0.00303$ & $0.50905 \pm 0.00385$ & $0.69229$ & $14.109$ \\
BSR10~\cite{bsr} & $1.05970 \pm 0.00093$ & $18.12348 \pm 0.17722$ & $0.10732 \pm 0.00312$ & $0.54362 \pm 0.00386$ & $0.72149$ & $13.891$\\ 
BSR15~\cite{bsr} & $1.05691 \pm 0.00088$ & $25.32453 \pm 0.23970$ & $0.09157 \pm 0.00296$ & $0.50424 \pm 0.00405$ & $0.67192$ & $14.523$ \\
BSR16~\cite{bsr} & $1.06169 \pm 0.00088$ & $24.13736 \pm 0.22708$ & $0.09312 \pm 0.00296$ & $0.49859 \pm 0.00399$ & $0.67167$ & $14.434$ \\
BSR17~\cite{bsr} & $1.06080 \pm 0.00090$ & $20.89719 \pm 0.19779$ & $0.09886 \pm 0.00302$ & $0.51927 \pm 0.00387$ & $0.68559$ & $14.311$ \\
FSUGZ00~\cite{fsugz} & $1.04892 \pm 0.00094$ & $23.00281 \pm 0.22990$ & $0.10562 \pm 0.00316$ & $0.57473 \pm 0.00435$ & $0.74736$ & $13.811$ \\
FSUGZ03~\cite{fsugz} & $1.07085 \pm 0.00090$ & $23.05967 \pm 0.21887$ & $0.09819 \pm 0.00302$ & $0.50820 \pm 0.00387$ & $0.69080$ & $14.108$ \\
FSUGZ06~\cite{fsugz} & $1.06429 \pm 0.00088$ & $24.51284 \pm 0.23012$ & $0.09214 \pm 0.00296$ & $0.49179 \pm 0.00403$ & $0.66915$ & $14.437$ \\
IUFSU$^*$~\cite{iufsus} & $1.04904 \pm 0.00096$ & $31.25275 \pm 0.33273$ & $0.10005 \pm 0.00322$ & $0.58418 \pm 0.00483$ & $0.78626$ & $14.319$ \\
G2$^*$~\cite{g2s} & $1.07406 \pm 0.00098$ & $28.57075 \pm 0.29748$ & $0.09629 \pm 0.00327$ & $0.57742 \pm 0.00497$ & $0.79847$ & $14.374$ \\
DD-F~\cite{ddf} & $1.05493 \pm 0.00092$ & $22.14656 \pm 0.21581$ & $0.10342 \pm 0.00309$ & $0.54042 \pm 0.00393$ & $0.71788$ & $15.238$ \\
DD-ME$\delta$~\cite{ddmedel} & $1.08153 \pm 0.00101$ & $19.87844 \pm 0.20471$ & $0.11910 \pm 0.00334$ & $0.56160 \pm 0.00391$ & $0.81266$ & $15.322$ \\
TW99~\cite{tw99} & $1.14237 \pm 0.00098$ & $19.81622 \pm 0.19257$ & $0.11633 \pm 0.00324$ & $0.46713 \pm 0.00361$ & $0.76282$ & $15.174$ \\
IUFSU~\cite{iufsu} & $1.22806 \pm 0.00093$ & $28.54012 \pm 0.26244$ & $0.09667 \pm 0.00306$ & $0.27626 \pm 0.00808$ & $0.70268$ & $14.489$\\
NL3~\cite{nl3}   & $1.12280 \pm 0.00104$ & $8.23918 \pm 0.10182$  & $0.12582 \pm 0.00341$ & $0.47125 \pm 0.00351$ & $0.82248$ & $14.549$\\
NL3$\omega\rho$~\cite{nl3wr} & $1.18772 \pm 0.00089$ & $31.72581 \pm 0.29654$ & $0.09049 \pm 0.00297$ & $0.27194 \pm 0.00850$ & $0.67947$ & $14.427$\\
\hline
\end{tabular}
\label{tab:fit}
\end{table*}

\begin{figure*}[!htb]
\centering
\includegraphics[scale=0.447]{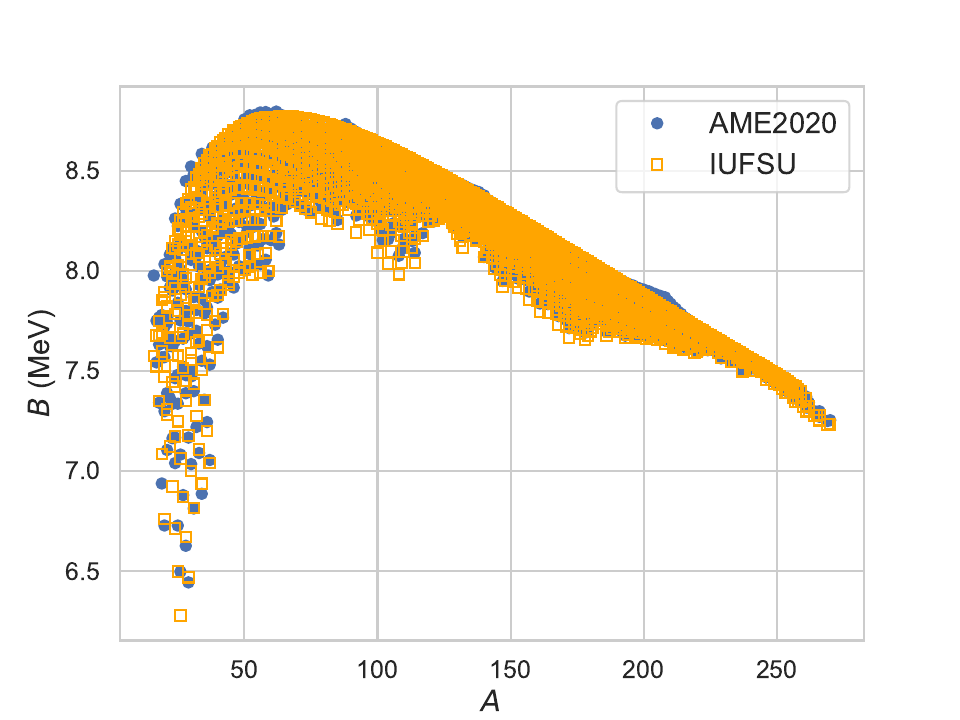}
\includegraphics[scale=0.447]{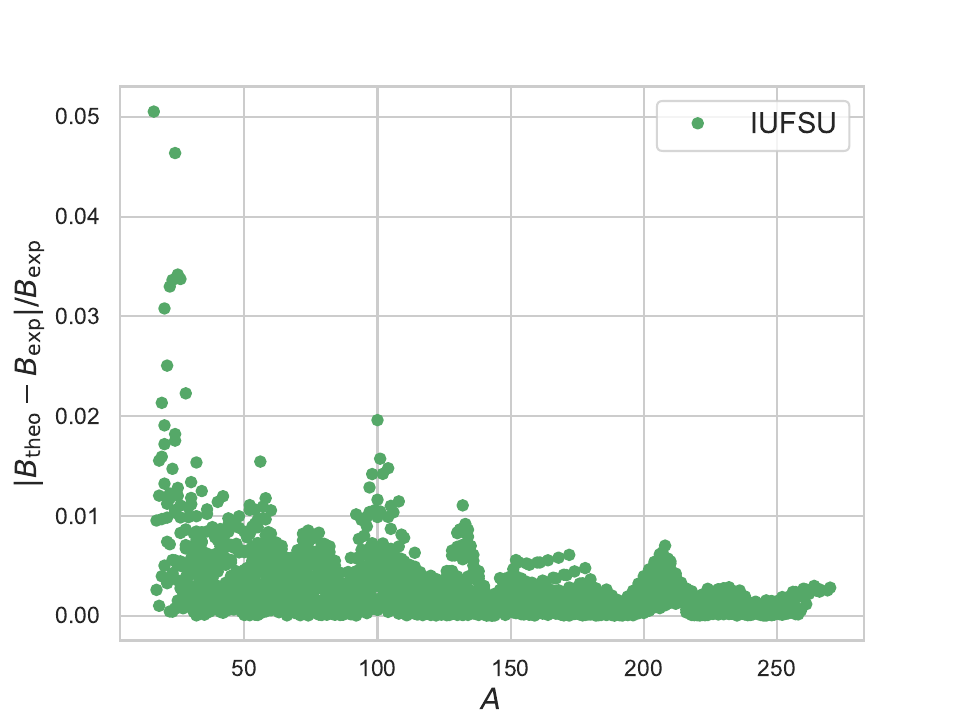}
\includegraphics[scale=0.447]{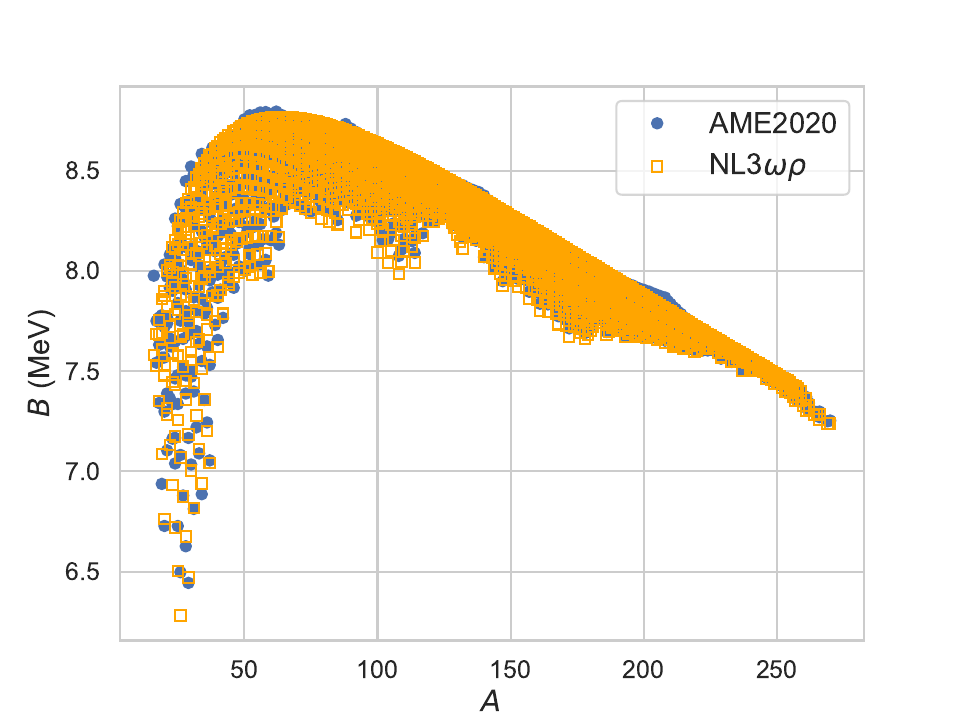}
\includegraphics[scale=0.447]{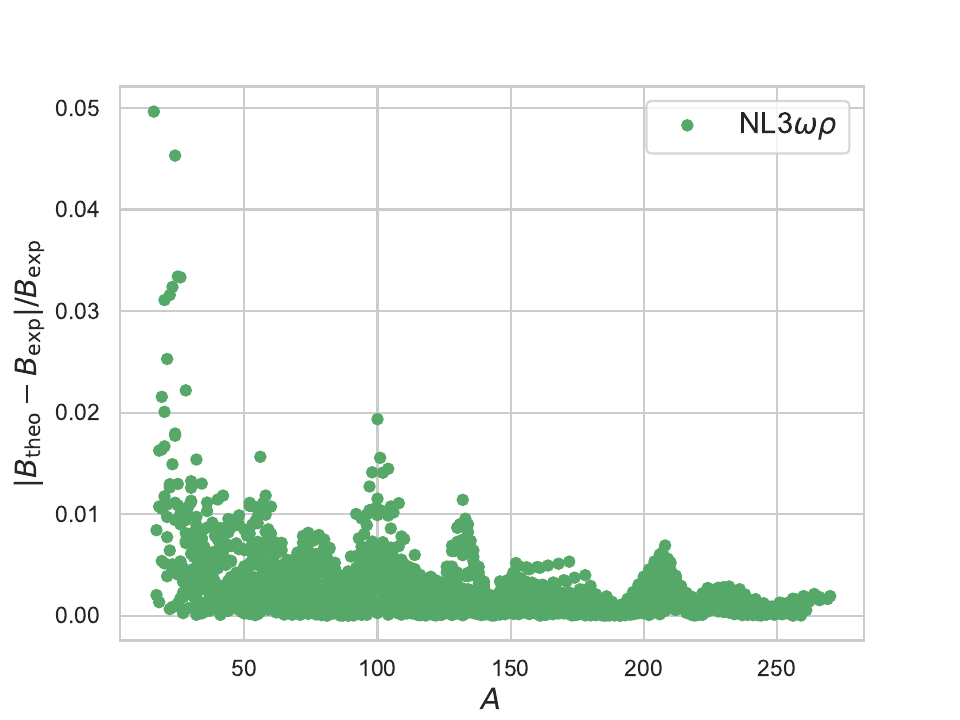}
\caption{Binding energy per nucleon calculated from IUFSU (left top panel) and \mbox{NL3$\omega\rho$} (left bottom panel) parametrizations compared to the respective experimental values taken from the AME2020. The relative errors in the binding energy per nucleon in the theoretical calculations are also shown for IUFSU (right top panel) and \mbox{NL3$\omega\rho$} (right bottom panel) parametrizations.}
\label{fig:fittings}
\end{figure*}

In Fig.~\ref{fig:fittings} it is depicted, for IUFSU and \mbox{NL3$\omega\rho$} parametrizations, the nuclear binding energy per nucleon (left panels), as well as the relative error of this quantity related to theoretical calculations and experimental values (right panels), all of them as a function of the mass number $A$.
As we one can verify from this figure, the experimental nuclear masses are very well reproduced, and the relative error is very low. The same pattern is observed for the remaining parametrizations used in this work.

\section{Surface and Curvature Tensions}

At finite temperature regime, surface and curvature tensions can be written as~\cite{Lattimer_1991,Hoa_2023}:
\begin{align}
\sigma_s(y,T) = \sigma_s(y,0)h(T),
\label{eq:sigmasT}\\
\sigma_c(y,T) = \sigma_c(y,0)h(T),
\label{eq:sigmacT}
\end{align}
where $\sigma_s(y,0)$ and $\sigma_c(y,0)$ are given in Eqs.~(\ref{eq:sigmast0}-\ref{eq:sigmact0}). The temperature dependence is encompassed by the function $h(T)$, given by
\begin{align}
h(T) = \left\{
\begin{array}{rcl}
0 \qquad \qquad  \qquad \mbox{if} \,\, T > T_c(y)\\
\left[1 - \dfrac{T^2}{T^2_c(y)}\right]^2 \,\, \mbox{if} \,\, T \le T_c(y)\\
\end{array}
\right.,
\label{eq:hT}
\end{align}
with the critical temperature at a particular proton fraction, $T_c(y)$, approximated by $T_c(y) = 4T_c^{\rm SM}y(1-y)$~\cite{Lattimer_1991}. Here, $T_c^{\rm SM}$ is the critical temperature related to symmetric matter (SM), i.e., at $y=0.5$, obtained from the solution of the coupled equations
\begin{align}
\frac{\partial P^{\rm SM}(n,T)}{\partial n}\Big|_{n=n_c^{\rm SM},T=T_c^{\rm SM}} = 0,
\\
\frac{\partial^2 P^{\rm SM}(n,T)}{\partial n^2}\Big|_{n=n_c^{\rm SM},T=T_c^{\rm SM}} = 0,
\label{eq:cond}
\end{align}
in which $P^{\rm SM}$ is the SM pressure and $n_c^{\rm SM}$ is the critical density. The values of $T_c^{\rm SM}$ of each parametrization used in this work is also provided in Table~\ref{tab:fit}.

\begin{figure}[!htb]
\centering
\includegraphics[scale=0.54]{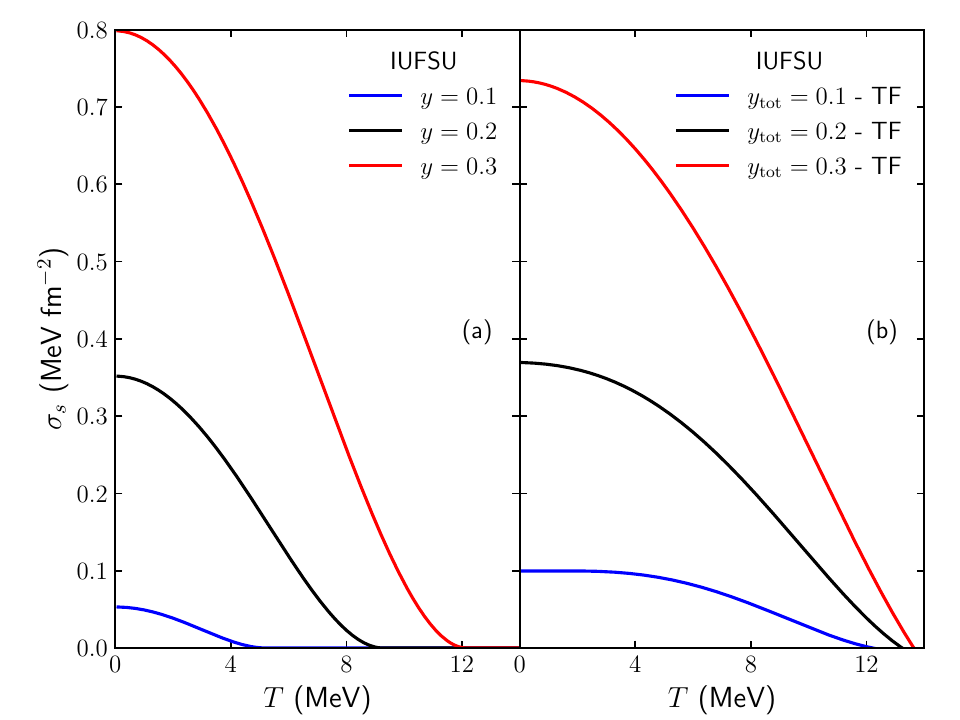}
\caption{Surface tension as a function of $T$ for IUFSU parametrization. Calculation (a) obtained through Eq.~(\ref{eq:sigmasT}) using the cluster proton fraction $y$ and (b) by the Thomas-Fermi approximation fitting using the total proton fraction $y_{\rm tot}$.}
\label{fig:sten-T-iufsu}
\end{figure}

\begin{figure}[!htb]
\centering
\includegraphics[scale=0.54]{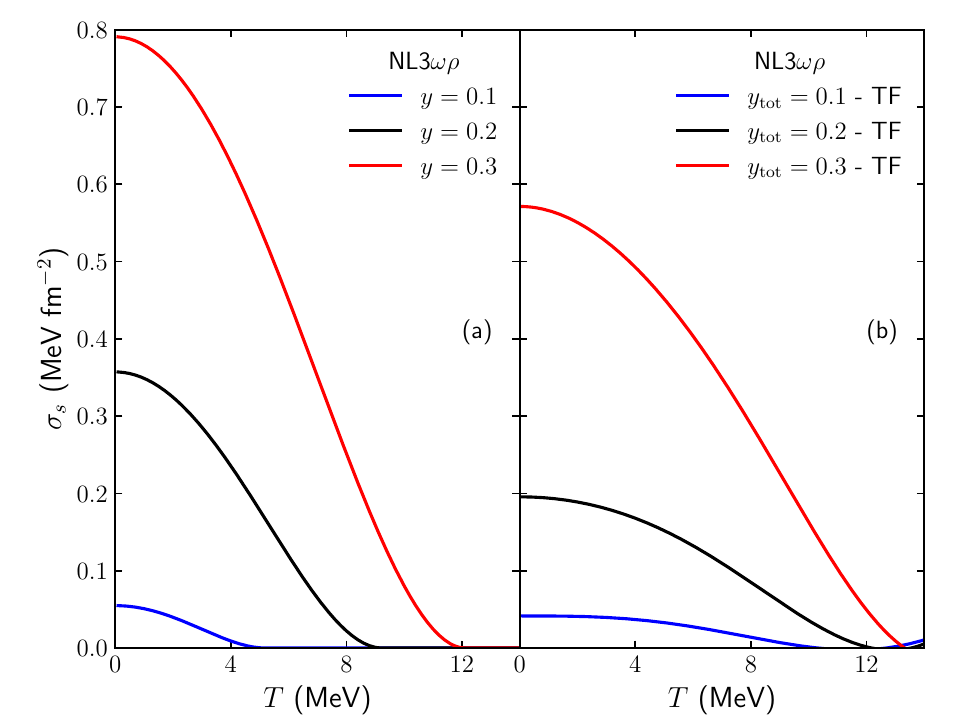}
\caption{The same as in Fig.~\ref{fig:sten-T-iufsu}, but for the NL3$\omega\rho$ parametrization.}
\label{fig:sten-T-nl3wr}
\end{figure}

In the following figures, we compare the results found from this approach to the ones obtained with the Thomas-Fermi approximation. In Figs. \ref{fig:sten-T-iufsu} and \ref{fig:sten-T-nl3wr} the surface tension is plotted as a function of the temperature respectively for the IUFSU and NL3$\omega\rho$ parameterizations. In Figs. (a), the surface tension is obtained with the prescription explained in the present work and the use of Eq. (\ref{eq:sigmasT}), where the proton fraction is the one of the denser phase, i.e., the cluster $y$. In Figs. (b), the surface tension is fitted by a Thomas-Fermi approximation and the proton fraction is the global one, $y_{tot}$. One can clearly see that an on-to-one mapping is not possible, except for $T=0$ when the total proton fraction is 0.5, which is exactly the cluster proton fraction.

\begin{figure}[!htb]
\centering
\includegraphics[scale=0.54]{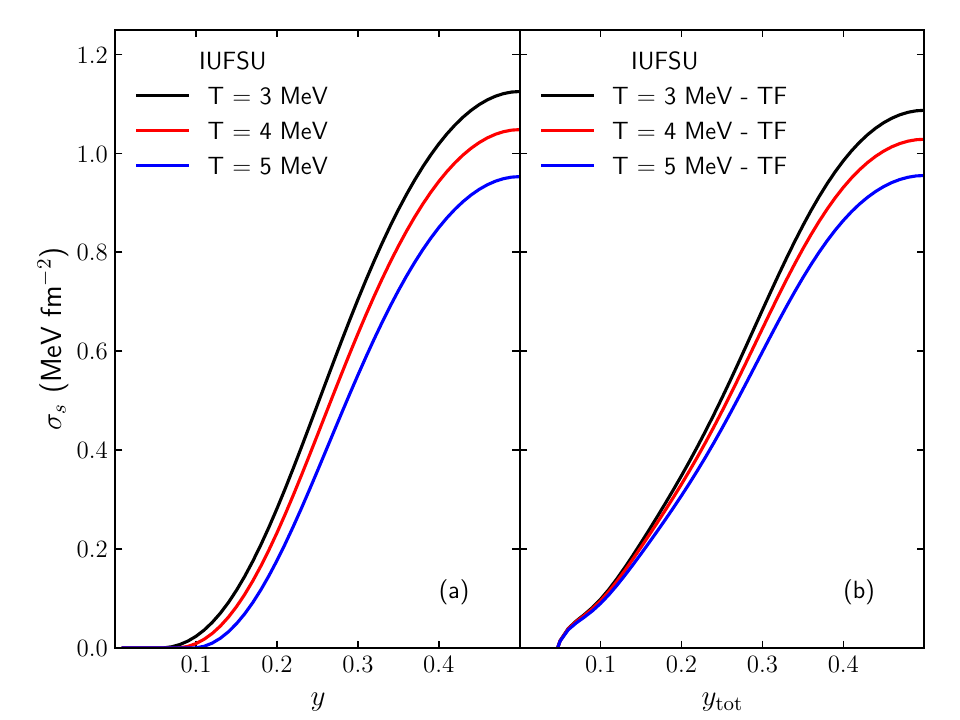}
\caption{Surface tension as a function of the respective proton fraction for IUFSU parametrization. Calculation obtained (a) through Eq.~(\ref{eq:sigmasT}) using the cluster proton fraction $y$ and (b) by the Thomas-Fermi approximation fitting using the total proton fraction $y_{\rm tot}$.}
\label{fig:sten-y-iufsu}
\end{figure}

\begin{figure}[!htb]
\centering
\includegraphics[scale=0.54]{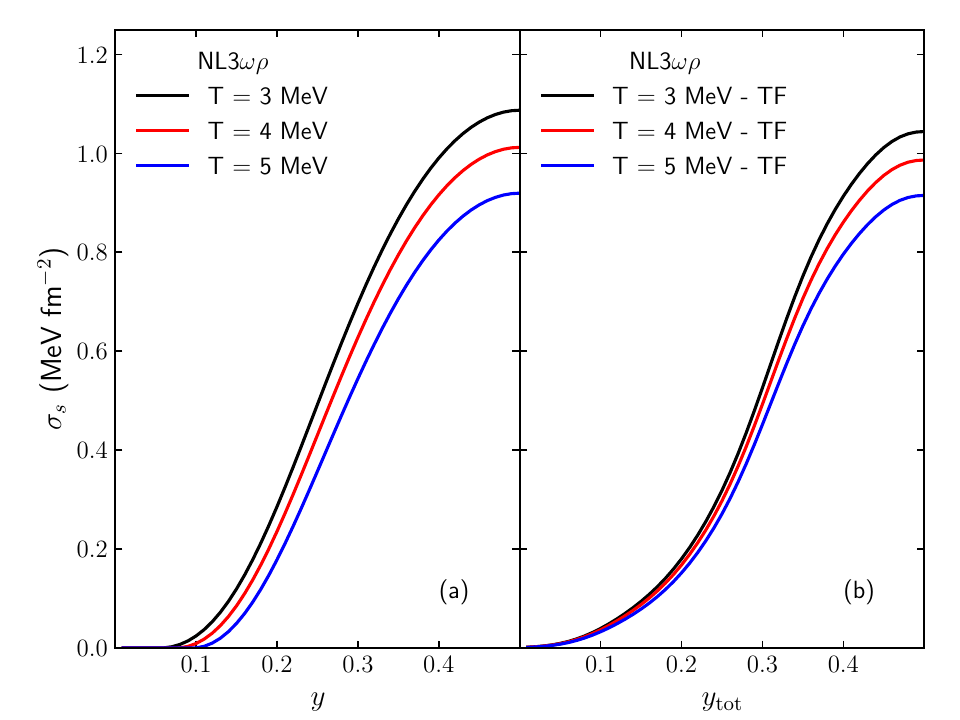}
\caption{The same as in Fig.~\ref{fig:sten-y-iufsu}, but for the NL3$\omega\rho$ parametrization.}
\label{fig:sten-y-nl3wr}
\end{figure}

In Figs. \ref{fig:sten-y-iufsu} and \ref{fig:sten-y-nl3wr}, the surface tension is displayed as a function of the proton fraction respectively for IUFSU and NL3$\omega\rho$ parametrizations. Again panels (a) and (b) refer, respectively, to the use of Eq.~(\ref{eq:sigmasT}) with the cluster proton fraction $y$ and the Thomas-Fermi approximation fitting using the total proton fraction $y_{\rm tot}$. One can observe that at the same temperature, both curves become more similar as the proton fraction approaches 0.5, but they are not identical, as the temperature imposes differences between the cluster and the total proton fraction. Only at $T=0$, there is no protons at all in the gas phase.  

\begin{figure}[!htb]
\centering
\includegraphics[scale=0.54]{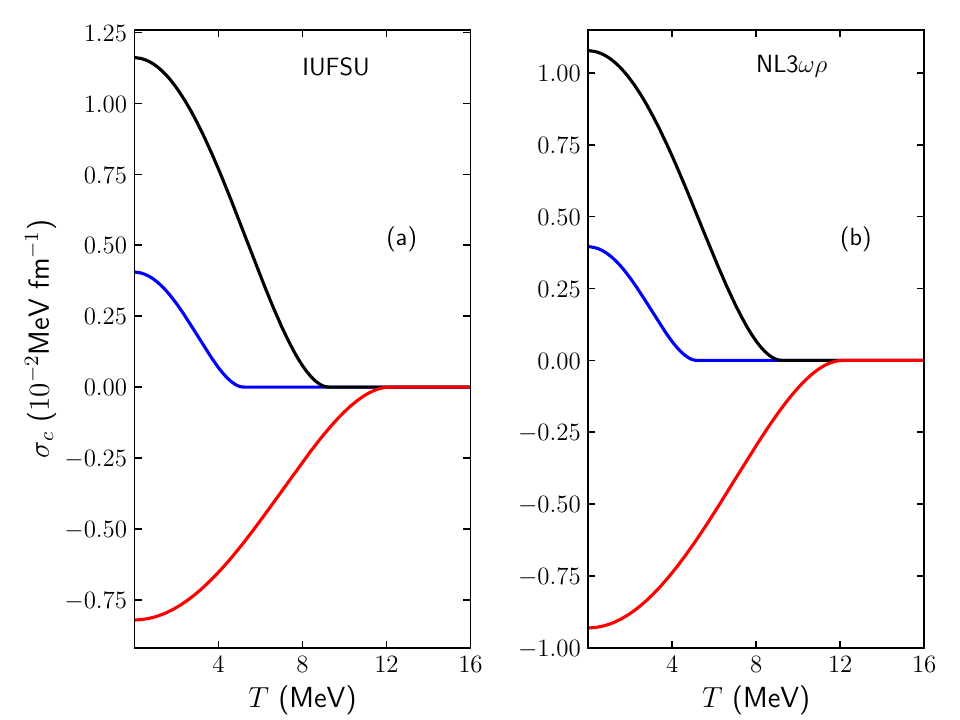}
\caption{Curvature tension as a function of $T$ obtained through Eq.~(\ref{eq:sigmacT}). Calculations for (a) IUFSU, and (b) NL3$\omega\rho$ parameterizations.}
\label{fig:cten-T}
\end{figure}

\begin{figure}[!htb]
\centering
\includegraphics[scale=0.54]{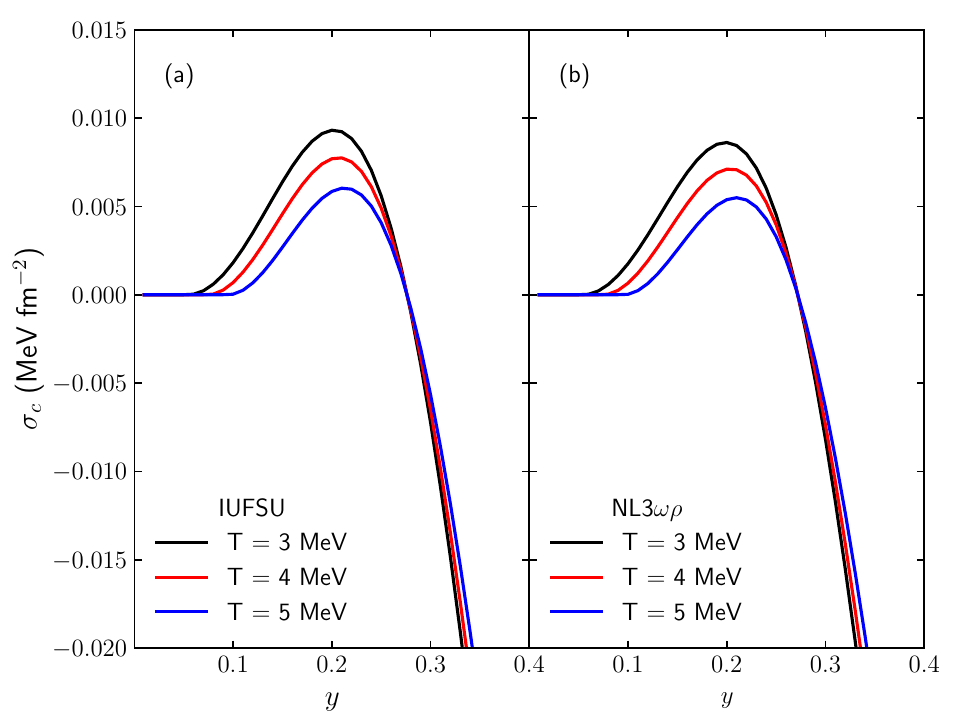}
\caption{Curvature tension as a function of $y$ obtained through Eq.~(\ref{eq:sigmacT}). Calculations for (a) IUFSU, and (b) NL3$\omega\rho$ parametrizations.}
\label{fig:cten-y}
\end{figure}

Finally, in Fig.\ref{fig:cten-T}, the curvature tension is shown as a function of $T$ obtained with the two parameterizations mentioned above and in Fig. \ref{fig:cten-y}, as a function of the cluster proton fraction. 

\begin{figure}[!htb]
\centering
\includegraphics[scale=0.54]{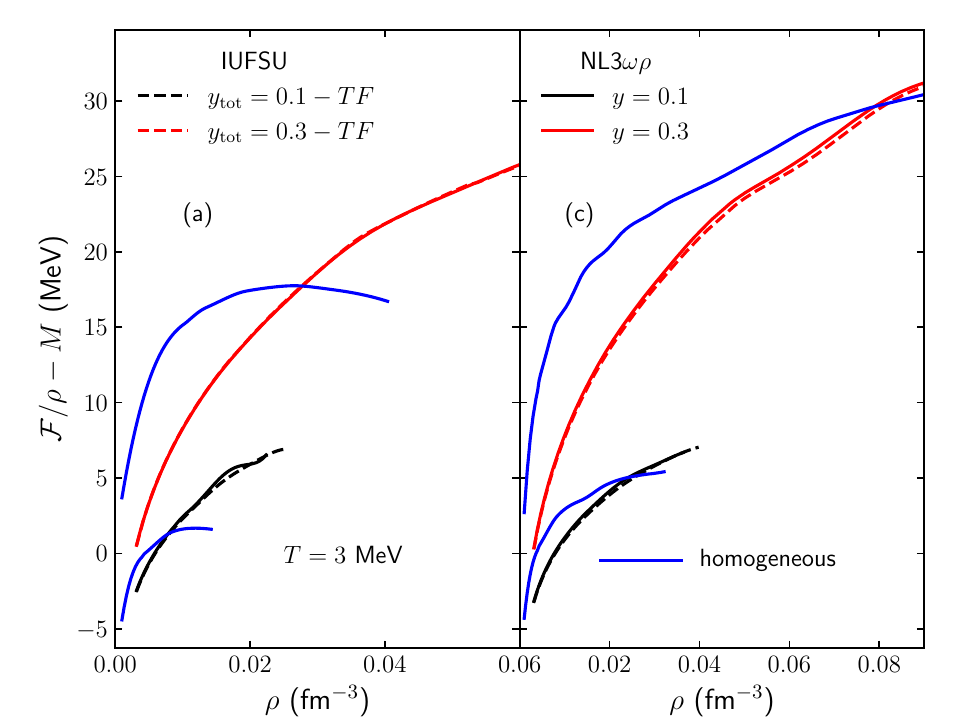}
\caption{Helmholtz free energy per particle versus density. Calculations at $T = 3$~MeV for (a) IUFSU and (b) NL3$\omega\rho$ parametrizations.}
\label{fig:comparison}
\end{figure}

At this point, we show that the pasta phase structure obtained with the 
prescription for the surface tension given in the present work and with the Thomas-Fermi approximation are indeed very similar, as seen in Fig. 
\ref{fig:comparison}{\color{blue}(a)} for IUFSU and
\ref{fig:comparison}{\color{blue}(b)} for NL3$\omega\rho$. 
The pasta phase was calculated with the CPA method already mentioned \cite{oldpasta,PRC82}, which assumes Gibbs conditions for phase coexistence. The homogeneous phase is also shown so that one can see that the transition from the pasta to the homogeneous phase takes place at either identical or very similar densities, confirming that the use of the fitting given in the present work is a very reasonable recipe for future calculations. Had we intended to correctly obtain the crust-core transition density in NSs, $\beta$-equilibrium and charge neutrality would have to be enforced at $T=0$. However, as we intend to check the results for proton fractions and temperatures of interest also to core-collapse supernova simulations, we have released these more strict NS conditions and
kept the calculations for fixed proton fractions and different temperatures. Nonetheless, the results for a proton fraction of the order of $0.1$ and low temperature are very close to the ones necessary to describe a NS.
In this case, the results obtained with both surface tension prescriptions are indeed very close and the crust-core transition density is coincident as shown.
For the sake of completeness, we also show the results for a larger proton fraction. One can see that the transition from the pasta to the homogeneous phase is the same, independently of the two surface tension parameterizations mentioned in this work.

\section{Final Remarks}

We have presented a simple prescription
to calculate parameterized expressions for the surface and curvature tensions based on the nuclear masses
of ionized nuclei and their binding energies. The pasta phase results obtained with the parametrized fittings were compared with the ones computed with the Thomas-Fermi fittings previously obtained for specific RMF models. Both free-energy densities are almost coincident and the transitions from the pasta phase to the homogeneous matter take place at practically the same densities. According to our calculations, the proposed prescription is simple, consistent, and quite robust and may be very useful when calculations involving surface and curvature tensions are necessary.

\section*{ACKNOWLEDGMENTS}
This work is a part of the project INCT-FNA proc. No. 464898/2014-5. It is also supported by Conselho Nacional de Desenvolvimento Cient\'ifico e Tecnol\'ogico (CNPq) under Grants No. 307255/2023-9 (O.L.), No. 308528/2021-2 (M.D.) and No. 303490/2021-7 (D.P.M.). O.~L. and M.~D. also thank CNPq for the project No. 401565/2023-8~(Universal). The authors gratefully thank Prof. Francesca Gulminelli for stimulating discussions and valuable comments.

%
\bibliography{references}{}
\bibliographystyle{apsrev4-2}

\end{document}